\documentclass{iau}
\usepackage{graphicx} 

\title[IAUS291.~~Unified equations of state of neutron stars and magnetars] 
{Unified description of dense matter in neutron stars and magnetars} 

\author[N. Chamel et al.]  
{N. Chamel$^1$, R. L. Pavlov$^2$, L. M. Mihailov$^3$, Ch. J. Velchev$^2$, Zh. K. Stoyanov$^2$, 
Y. D. Mutafchieva$^2$, M. D. Ivanovich$^2$, A.F. Fantina$^1$,J.M. Pearson$^4$ \and S. Goriely$^1$}

\affiliation{$^1$Institute of Astronomy and Astrophysics, Universit\'e Libre de Bruxelles, CP 226, Boulevard du Triomphe, B-1050 Brussels, 
Belgium\\[\affilskip]
$^2$Institute for Nuclear Research and Nuclear Energy, Bulgarian Academy of Sciences, 72 Tsarigradsko Chaussee, 1784 Sofia, Bulgaria\\[\affilskip]
$^3$Institute of Solid State Physics, Bulgarian Academy of Sciences, 72 Tsarigradsko Chaussee, 1784 Sofia, Bulgaria\\[\affilskip]
$^4$D\'ept. de Physique, Universit\'e de Montr\'eal, Montr\'eal (Qu\'ebec), H3C 3J7 Canada}

\pubyear{2012}
\volume{291}  
\jname{\mbox{Neutron Stars and Pulsars: Challenges and Opportunities after 80 years}}
\editors{J. van Leeuwen, ed.} 
\begin{document}

\maketitle

\begin{abstract}
We have recently developed a set of equations of state based on the nuclear energy density functional theory providing 
a unified description of the different regions constituting the interior of neutron stars and magnetars. The nuclear 
functionals, which were constructed from generalized Skyrme effective nucleon-nucleon interactions, yield not only an excellent fit to essentially all 
experimental atomic mass data but were also constrained to reproduce the neutron-matter equation of state as obtained from 
realistic many-body calculations. 
\end{abstract}


\firstsection 
\section{Introduction}

With a mass of the order of that of our Sun compressed inside a radius of about $10$~km only, 
neutron stars (NS) are among the most compact objects in the universe (see e.g. \cite{haensel2007}). Their central 
density can exceed several times the density encountered in the heaviest atomic nuclei. NS are 
also the most strongly magnetized objects. Surface magnetic fields of order $10^{14}-10^{15}$~G have 
been estimated in soft gamma-ray repeaters and anomalous X-ray pulsars assuming that their spin-down is 
due to magnetic dipole radiation (see e.g. \cite{mcgill}). In addition, circumstantial evidence of surface 
magnetic fields greater than $10^{15}$~G have been reported from spectroscopic studies (see e.g. \cite{stro2000,gavr2002,woods2005}).
The internal magnetic field of a NS could be even stronger than its surface field, as found in the Sun (see e.g. \cite{solanki2006}). 
In particular, according to the magnetar model of \cite{td93}, magnetic fields up to $\sim 10^{17}$~G could be generated via dynamo 
effects in hot newly-born NS. 

The outer crust of a cold non-accreting NS is primarily composed of pressure ionized iron atoms arranged in a regular crystal 
lattice and embedded in a highly degenerate electron gas. With increasing density, nuclei become more and 
more neutron-rich due to electron captures. Eventually, at a density $\sim 4 \times10^{11}$ g/cm$^{3}$, some 
neutrons start to drip out of nuclei, thus defining the boundary between the outer crust and the inner crust. At densities 
above $\sim 10^{14}$~g/cm$^3$, the crust dissolves into a uniform plasma of nucleons and leptons. The composition of 
the core remains very uncertain. 

We have determined the internal structure a cold non-accreting NS endowed with a strong magnetic field using a unified
treatment of dense matter based on the nuclear energy-density functional (EDF) theory.

\section{Brussels-Montreal equations of state}

The EDF theory provides a self-consistent description of various nuclear systems, from finite nuclei to homogeneous 
nuclear matter. It is therefore well suited for studying the interior of a NS. 
The Brussels-Montreal EDF BSk19, BSk20 and BSk21 were derived from generalized Skyrme effective nucleon-nucleon interactions 
which fit essentially all measured masses of atomic nuclei (calculated using the Hartree-Fock-Bogoliubov method) with an 
rms deviation as low as 0.58 MeV. In addition, these EDF were constrained to reproduce three different representative 
neutron-matter equations of states (EoSs) obtained from microscopic calculations using realistic two- and three- body forces 
and reflecting the current lack of knowledge of dense neutron matter (see \cite{goriely2010}). 

We used these EDF to calculate consistently the properties of all regions of the interior of a non-accreting NS, from its 
surface down to the center, under the assumption of cold catalyzed matter (see \cite{pearson2011, pearson2012}). The core was assumed 
to be made of nucleons and leptons only. The resulting unified EoSs are consistent with the radius constraints of \cite{ste10} 
inferred from observations of X-ray bursters and low-mass X-ray binaries (~\cite{fantina2011}). However, only the EoSs based on 
the BSk20 and BSk21 EDF are stiff enough at high densities to support NS as massive as PSR J1614$-$2230 (see \cite{chamel2011}). 

\section{Internal structure of magnetars}

The internal composition of a magnetar can be substantially different from that of an ordinary NS, especially in the outermost layers. 
We have therefore recalculated the EoS of the outer crust of a NS taking into account the presence of the magnetic field using 
the BSk21 EDF (see \cite{chamel2012}). In a strong magnetic field, the electron motion perpendicular to the field is quantized into 
Landau levels and this can change the sequence of equilibrium nuclides. We have found that the deviations become particularly significant for 
$B\sim 10^{16}$~G. Moreover, strong magnetic fields tend to prevent neutrons from dripping out of nuclei thus increasing the pressure at 
which the neutron drip transition occurs.
The effects of the magnetic field on nuclei, which we have neglected, could also have an impact on the crust for 
$B\gtrsim 10^{17}$~G (\cite{peana2011}).

Strong magnetic fields can change the EoS in the surface regions where only a few Landau levels are filled. 
However, with increasing density the effects of $B$ become less and less important as more and more levels are populated and the EoS matches 
smoothly with that obtained for $B=0$. For $B\sim 10^{15}$~G, only the EoS in the outer crust is affected. Therefore the global structure of a 
magnetar would be almost undistinguishable from that of an ordinary NS.

\section{Conclusions}

We have developed a series of EoSs of cold catalyzed matter based on the EDF theory and describing consistently all regions of a cold 
non-accreting NS (\cite{chamel2011}). These EoSs have been recently extended to magnetars by taking into account the effects of the strong 
magnetic field in the outer crust (\cite{chamel2012}). We have found that the outer crust of a magnetar could have a substantially different 
composition (hence also different properties) compared to that of an ordinary NS. 

\begin{table}
\centering
\caption{Sequence of equilibrium nuclides with increasing depth in the outer crust of a cold non-accreting magnetar endowed with a magnetic field $B=10^{17}$~G. For comparison, the results obtained for $B=0$ are also indicated. }
\label{tab1}
\vspace{.5cm}
\begin{tabular}{|cc|}
\hline
  $B=10^{17}$~G & $B=0$  \\
\hline
$^{56}$Fe & $^{56}$Fe \\
$^{62}$Ni & $^{62}$Ni \\
 - & $^{58}$Fe \\
- & $^{64}$Ni\\
- &   $^{66}$Ni\\
$^{88}$Sr & - \\
$^{86}$Kr & $^{86}$Kr\\
$^{84}$Se &  $^{84}$Se\\
$^{82}$Ge & $^{82}$Ge\\
$^{132}$Sn & -\\
$^{130}$Cd & -\\
$^{128}$Pd & - \\
$^{126}$Ru & - \\
- & $^{80}$Zn\\       
-  & $^{79}$Cu  \\
- & $^{78}$Ni  \\
-  & $^{80}$Ni  \\
$^{124}$Mo &$^{124}$Mo  \\
$^{122}$Zr & $^{122}$Zr \\
$^{121}$Y  &  $^{121}$Y \\
$^{120}$Sr &  $^{120}$Sr \\
$^{122}$Sr & $^{122}$Sr \\
$^{124}$Sr &  $^{124}$Sr\\
\hline             
\end{tabular}    
\end{table}

\acknowledgments This work was supported by FNRS (Belgium), NSERC (Canada), Wallonie-Bruxelles-International (Belgium), the 
Bulgarian Academy of Sciences and CompStar.

\end{document}